# Swarm coordination of mini-UAVs for target search using imperfect sensors

Antonio L. Alfeo[1], Mario G. C. A. Cimino[1,*], Nicoletta De Francesco[1], Alessandro Lazzeri[1], Massimiliano Lega[2], Gigliola Vaglini[1]

[1] Department of Information Engineering, University of Pisa, Largo L. Lazzarino 1, 56127 Pisa, Italy

[2] Department of Engineering, University of Naples "Parthenope", 80143 Naples, Italy

*luca.alfeo@ing.unipi.it, mario.cimino@unipi.it, nicoletta.defrancesco@unipi.it, alessandro.lazzeri@for.unipi.it, lega@uniparthenope.it, gigliola.vaglini@unipi.it*

\* Corresponding author:

Mario G. C. A. Cimino, mario.cimino@unipi.it

Tel: +39 050 2217 455; Fax: +39 050 2217 600.

**Abstract** Unmanned Aerial Vehicles (UAVs) have a great potential to support search tasks in unstructured environments. Small, lightweight, low speed and agile UAVs, such as multirotors platforms can incorporate many kinds of sensors that are suitable for detecting object of interests in cluttered outdoor areas. However, due to their limited endurance, moderate computing power, and imperfect sensing, mini-UAVs should be into groups using swarm coordination algorithms to perform tasks in a scalable, reliable and robust manner. In this paper a biologically-inspired mechanisms is adopted to coordinate drones performing target search with imperfect sensors. In essence, coordination can be achieved by combining stigmergic and flocking behaviors. Stigmergy occurs when a drone releases digital pheromone upon sensing of a potential target. Such pheromones can be aggregated and diffused between flocking drones, creating a spatiotemporal attractive potential field. Flocking occurs, as an emergent effect of alignment, separation and cohesion, where drones self organise with similar heading and dynamic arrangement as a group. The emergent coordination of drones relies on the alignment of stigmergy and flocking strategies. This paper reports on the design of the novel swarming algorithm, reviewing different strategies and measuring their performance on a number of synthetic and real-world scenarios.

## 1 Introduction and Problem Statement

In recent years, several research groups are working on new procedures and technologies to operate and monitor complex scenarios. Two specific areas include search and rescue and environmental monitoring. Both these topics require solutions to critical issues related to the mission requirements and the mission profile. The choice of a specific aerial platform for the monitoring of complex scenarios should be made by examining particular correspondence to the needs of the mission at the same time, and the multiplying effect of what is measurable by sensors positioned on the ground as fixed configuration.

Advanced aerial platforms such as Unmanned Aerial Vehicles (UAVs), often called drones, are today the most frequent response to the needs of different missions. In particular, the specific category of mini-UAV is the perfect "solution" for the missions that are generally categorized as the "three Ds" (dull, dirty, or dangerous). Moreover, drones have recently received a strong technological acceleration thanks to recent advances in miniaturization of battery, of communication, processing and sensing technology [1].

The remote/proximal sensing data obtained using mini-UAVs were validated in several environmental monitoring missions with complex scenarios as reported in previous research; these include: fusion of optical data with synthetic aperture radar data to detect environmental hazards [2,3], use of thermal imagery to monitor landfills [4], surface waters contamination [5] and to detect illegal dumping [6,7] and to identify other illegal activities [8]. In addition, remote sensing data can be strategically combined with other data layers in geographic information systems to monitor the vulnerability of cultural sites [9] and anticipate environmental violations [10,11].

The use of a range of aerial platforms and advanced sensors to detect the illegal activities was validated in several real missions in Italy [12,13]. These are the first known use of these methods in both the fields of environmental research and law enforcement/environmental forensics. They also represent an example of collaboration between law enforcement and university teams on developing enhanced environmental protection methods.

In the operational surveillance for successful identification and prosecution of environmental pollution culprits it is required an integrated system based on data from several sources. The surveillance service must also include geospatially tagged forensic data analysis (information arising from navigation/positional systems).

The detection, identification and localisation of a target are key elements in all the above operations. Groups of mini-UAVs equipped with self-localisation and sensing capabilities offer new opportunities; indeed, groups of mini-UAVs can explore cluttered outdoor environments, where access to conventional platforms is inefficient, limited, impossible, or dangerous. In brief, the main motivations for adopting the UAV technology in the survey process are the following: reduction of risk of human falling, reduction of safety costs for plant stoppage, improved data density and quality due to a better proximity, accessibility to locations where people or vehicles have no access, faster and cheaper data acquisition due to the involvement of less workforce and equipment.

The coordinated swarming drones could be also considered as single array of sensors configured to the measure of a host of environmental parameters. In search and rescue tasks, for example, a more effective approach is to achieve a quick "survey" of the area to identify key locations as quick as possible. This exclusion process enables organisers to rescan the key locations that provided some circumstantial evidence. In this context, the quality of the sensing has also a direct impact on the overall mission performance [14].

Therefore, an important aspect of the swarm coordination is the possibility to require a sufficient number of redundant samples of the target to reliably classify it as "detected" or "undetected".

A cooperative approach that exploits drones sensing, minimizes the error in target recognition [15]. In contrast, to use a unique drone implies costly structure and design, as well as vulnerability. Hence, a number of considerations support the use of coordinated swarming drones. An important requirement of the coordination strategy is to avoid centralized control approaches, leading to exponential increases in communication bandwidth and software complexity [16]. Swarm intelligence methodologies can be investigated to solve problems cooperatively while maintaining scalability. The main inspiration for swarming drones comes from the observation of social animals, such as insects, winged animals, and fish, that exhibit a collective intelligence which appears to achieve complex goal through simple rules and local interactions [17]. The main benefits of a swarm drones includes: robustness (for the ability to cope with the loss of individuals); scalability (due to the ability to perform well with different group size); and flexibility (thanks to the capability to manage a broad spectrum of different environments and tasks). To this aim, each individual of the swarm: acts with a certain level of autonomy; performs only local sensing and communication; operates without centralized control or global knowledge, and cooperates to achieve a global task [17].

In this paper, different coordination strategies are reviewed and tested empirically with both synthetic and real-world scenarios, with obstacles having irregular complex shapes. For this purpose, it is adopted a multi-agent simulation platform with the possibility of importing environments with obstacles and targets sampled from real landscapes.

The paper is structured as follows. In Section 2 early requirements and coordination strategies are reported. Section 3 briefly characterizes the related work. In Section 4, the analysis and the integration of the emergent schemes is covered. Section 5 reports on the design of the algorithm. Experimental studies are detailed in Section 6. Section 7 draws conclusions and future work.

## 2 Early requirements and coordination strategies

From a structural standpoint, it is assumed that each mini-UAV is provided with the following capabilities: wireless communication capability for sending and receiving information from the ground station; self-location capability based on Global Position System (GPS) and inertial technology, returning the coordinates of its current location; one or more target sensing technology, capable of acquire data in the area over which it flies; processor with limited computing capability; obstacle avoidance capability, that is, locally managed detection and steering to avoid flying towards surrounding barriers and drones. Moreover, it is assumed that a certain level of uncertainty comes from noisy of faulty sensor measurements.

Marker-based stigmergy is a fundamental swarm coordination mechanism, based on the release of information in the environment in the form of pheromones [18,19]. The pheromone is a volatile substance diffused locally and staying temporarily for other individuals that can properly react and modify their behavior [20]. Simulated (that is, digital) pheromones can be used to coordinate groups of drones for various tasks. In a distributed environment, a pheromone map of the search space can be maintained and made available for drones as a "remote brain" capability [21].

When the sensing system of a drone determines a potential target, it tries to trigger the cooperation of its swarm to achieve *reliable sensing* and *target detection*.

- *Reliable sensing*. Sensors on mini-UAV can generate faulty measurements for a number of reasons, such as power loss, software failure, small bias, miscalibration, slow-drifting, loss of accuracy, temporary freezing, to name a few [22]. In the literature, some approaches to fault recognition assume the fault types can be described by a static parameterized model. If parameterized models for the fault types are available, a fault recognition algorithm can be applied. However, without health monitoring a static fault model is often not known [22]. In practice, there is likely to be some deviation between what the actual faults look like and what their models predict. This residual may be irrelevant for a single sensor system. However, in a conventional distributed approach even small residuals can have a significant impact on the overall effectiveness, due to the high number of occurrences potentially involved. In contrast, swarm systems could exhibit much higher levels of robustness, in the sense of tolerance to individual (or few) residual(s), than in conventional distributed systems. Nevertheless, a simplistic modelling approach may make incorrect assumptions, because the question of how many agents are needed to guarantee a required emergent behavior in a particular swarm and for a particular behavior is not straightforward [23]. This *potential* tolerance cannot be natively assumed without special analysis, design, and test, since swarm systems can exhibit a number of unexpected behaviors. Therefore, the proposed drones' coordination algorithm needs to incorporate some mechanism able to exploits the inherent collective influence between measures, in order to verify its effectiveness under assumption of uncertainty in individual sensing. To this aim, this study tries to achieve a control on the number of redundant measures of the targets that are sufficient to ensure a sufficient level of reliability.

- *Target detection*. For a distributed target, the detection process is the identification of any parts of it, with sufficient detail to permit the intended action. For example: to detect a landmine means to find the location of it to avoid being maimed or killed. To detect radioactive substance means to trace perimeters were radioactivity levels are considered dangerous. To detect gas leak means to identify the area were natural gas seeping from the ground implies fire and explosion hazards, and so on. The search problem is formulated by discretizing the environment into a set of cells. Each target is stationary and usually covers many cells. The objective is to determine in which cells the targets lie. Due to the distribution, the task requires that drones are dynamically arranged so as to be efficiently engaged when some member detects a part of the possible target.

For this purpose, the drone releases a particular amount of pheromone on the cell of the sensed possible target, whose diffusion acts as an attractive potential on neighboring drones. To be attracted by pheromone trails, the available drones should be spatially organized into flocks. Flocking is a strategy to allow the self-organization of drones into a number of flocks. Flocking behavior is an emergent effect of individual rules based on alignment, separation and cohesion [24]. With *alignment* rules the drones tends to move in the same direction that nearby drones. With *separation* rules, the drone keeps a minimum distance able to provide the drone with flexibility when moving in the swarm, and for a better exploration. Finally, with *cohesion* rules the drone tends to move towards the swarm.

As a result of flocking, each member of a flock has approximately the same heading of the other members, and attempts to remain in range with them. For this purpose, the structural dimensions of the pheromone should take into account the average size of a swarm (or vice versa). Otherwise, a highly diffused or poorly evaporated pheromone could attract disproportionate resources on a single target, thus interfering with the progressive development of the emergent behavior. In contrast, a poorly diffused or highly evaporated pheromone could not be sensed at all.

As an effect of pheromone attraction, other drones can confirm the possible target through repeated sensing, and can surround the detected location in order to map the whole distribution. Thus a considerable amount of pheromone is aggregated for each possible target. Once a predefined number of drones confirmed the sensing of the possible target, it is definitively considered to be a true target and then its sensing cannot activate additional pheromone. Since pheromone evaporates over time, after a certain time the pheromone intensity cannot be reinforced in a fully explored region, and in practice disappears.

In the presented approach, stigmergy and flocking are two emergent behavioral patterns which should work in conjunction with other basic behavioral patterns of the drone, such as obstacle and boundary avoidance. The process of designing a combination strategy is bottom-up and consists in finding the right setting at the micro-level (agent-level) in order to obtain a coherent emergent behavior at macro-level (swarm-level) [25].

## 3 Related Work

The goal of this section is to briefly characterize the main approaches and results in the literature on stigmergic mechanisms coordinating swarms of small robots to perform target search or similar tasks. The published works in the field can be distinguished into three categories: using a physical substance as a pheromone, which is necessarily transmitted in an indirect way between robots, by means of the physical environment; using a digital pheromone, transmitted via direct communication between robots; using a digital pheromone, transmitted via an indirect communication between robots. The latter is the category of our approach.

Kuyucu *et al.* in [26] use a swarm of robots releasing physical substance as a repulsive pheromone, for environment exploration. In particular, robots act combining three basic behaviors, with decreasing priority: wall avoiding, pheromone coordination, and random walk. Actually there are various approaches in the literature using physical pheromones, because they do not require a computational structure. Although real pheromones are not usable with aerial vehicle, they can be simulated. Thus, this type of research can be interesting to model new types of digital stigmergy.

An example of stigmergic coordination between drones using direct communication is presented by Dasgupta [27], where he focuses on automatic target recognition. Potential target are marked by drones, which also communicate the gossiped pheromone to nearby drones, with probability inversely proportional to the distance from the source. The proposed stigmergic schema employs also repulsive pheromone, as a negative feedback, when a predefined number of drones identify the same target. A disadvantage of such scheme is that the bandwidth required goes into an exponential explosion as the population grows. To avoid redundancy in target evaluation each UAV has to maintain in memory the state of each potential and confirmed target. In this way, the direct communication in the swarm should be strongly limited [28].

A swarm coordination schema with indirect coordination is proposed by Sauter *et al.* [18]. Here the coordination of a swarm of vehicles is based on digital pheromones maintained in an artificial space called pheromone map and composed by an arbitrary graph of place agents, that is, intermediate control nodes. There are two classes of agents which deposit, withdraw, and read pheromones, that is, walkers and avatars. A walker agent aims to make movements and action decisions, whereas avatars collect location information to make estimates when sensor information is not available. The schema has been applied to a range of scenarios, among which target acquisition. An important problem of this approach is that the exploration depends on the initial position of the swarm. This model does not consider complex targets but only simple targets without structure.

To handle the unreliability in sensing, a certain number of drones must be attracted on a potential target. To achieve this goal a spatial organization of the available drones is required in order to sense the pheromone deposit released during a survey leaded by a peer of the same group. This result can be achieved keeping flocking formation. Flocking behavior is exhibited during the birds' group flight. It is an emergent effect caused by the observance of three rules: preserving heading alignment with flock-mates, while maintaining separation with respect to the nearest one and cohesion with the entire group, as described by Reynolds [24]. This flocking behavior formalization have been extensively used in swarming robots and drones coordination. Bouraqadi *et al.* [29] accomplish an unknown environment survey via a group of robots which has to stay close enough to maintaining the ability to communicate with each other. This objective is reached using Reynolds rules to organize the robots distribution and movements. Hauert *et al.* [30] apply flocking rules for the management of a drones swarm in order to keep an ad-hoc network during their flight and to coordinate their task. However, this application is based on the assumption of well-known search field, and then it is not applicable to unstructured environments, which is one of our requirements.

## 4 Behavioral specification of the proposed approach

This section aims to characterize the emergent behavior of the coordination algorithm, via the integration of a variety of mechanisms. This purpose is achieved using the Tropos agent-oriented methodology [31]. Tropos is based on the notion of *agent*, which in this context is a drone, with related notions such as *goals* and *plans*. It allows a clear modeling of the operating environment and of the interactions that should occur between drones. Figure 1a shows a legend of the main concepts: *actor*, *goal*, *plan*, *resource*, *capability*, and *social dependency* between actors for goal achievement. Actors may be further specialized based on *roles* (circle with a bottom line) or agents (circle with upper line). A software agent represents a physical instance (human, hardware or software) of an actor that performs the assigned activities. A role represents a specific function that, in different circumstances, may be played by the agents. Edges Edges between nodes form dependencies of the form: "actor → goal/task/resource → actor". In additional to hard goals, soft goals are also used when having no clear-cut definition and/or criteria as to whether they are satisfied, for example for modeling goal/plan qualities and non-functional requirements [32]. A detailed account of modeling activities can be found in [31].

Figure 1c represents a top view of the proposed algorithm. More specifically, on the bottom, a *Physical Environment* is a resource modeling the search field, which contains all the physical elements interacting with drones, whereas a *Virtual Environment* is a resource managing virtual pheromones and the targets (cells discovered or not). In the middle, *Drone* is the main actor, supporting the primary goal *look for target*, collectively attained by two levels of organization: the *flock*, that is, the organism consisting in locally coordinated *drones*, and the *swarm*, that is, the organism consisting in globally coordinated *flocks*. Conversely, a *drone* depends on the *swarm* for *saving fly time*, since the coordinated search is purposely organized to reduce the overall time. This purpose is based on the resource *accomplishment time*, managed via *update* plans of the *virtual environment* (on the bottom right of the figure): *count target found* and *time unit*. Other update plans of the *virtual environment* are *diffuse and evaporate pheromones*. The basic needs of a *drone* consist in the *sensing procedures*, carried out via both the *physical* and *virtual environments*, whereas the basic *soft goals* of a *drone* consist of: *to cover the search space, collective flight, to point towards targets*, and *to follow obstacle-free paths*. Such soft goals are attained via related roles (in ascending order of priority): *obstacle avoider*, *tracker*, *flockmate*, and *explorer* [33]. Above all, Fig. 1d shows how a drone *reacts to local conditions*. Each role is further detailed in Fig. 1b. Fig. 1e represents the *obstacle avoider* role, with the first priority level. At the second priority level, Fig. 1f represents the *tracker* role. Fig. 1b shows the third priority level, *flockmate*. Finally, Fig. 1g represents the minimum priority level, *explorer.*

The above specification is a mixed actor-dependency model in which dependencies/delegations among emergent actor are highlighted while agents' behavior is explained. As a result, the drone task and goals and its precedencies between roles have been detailed. The next Section focuses on the system design, to show how to implement and integrate the main functional and architectural components.

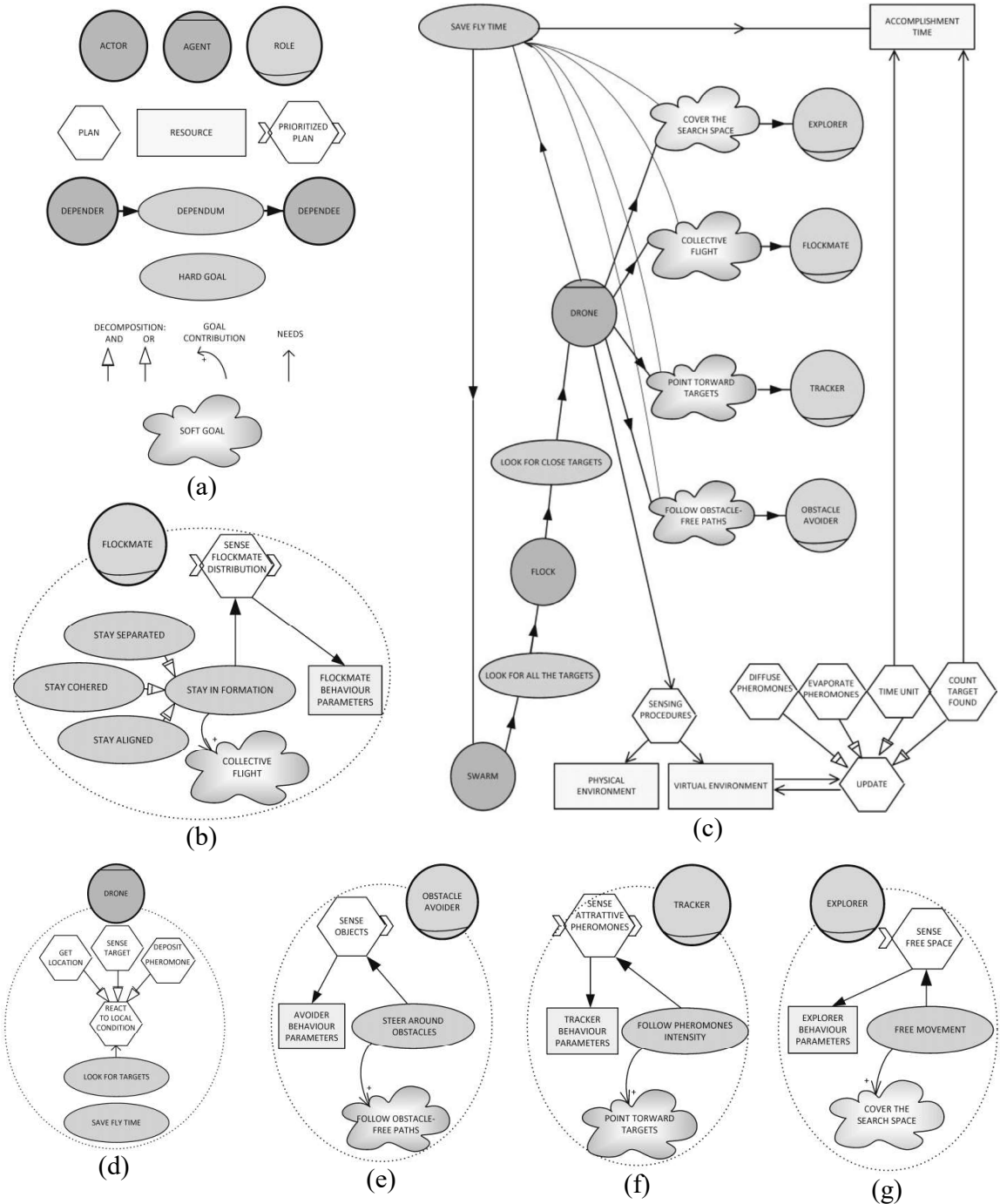

**Fig. 1.** Behavioral representation of the proposed approach

## 5 Architectural and functional Design of the main subsystems

This section is devoted to the modeling of environment and drones.

*4.1 The design of the environment: pheromones and error dynamics*

It is assumed that the environment is constrained to a specific area. Without loss of generality, the area is discretized through a grid consisting of $C^2$ cells, each identified by a pair *(x,y)* of coordinates, with $x,y \in [1,\dots,C]$. The actual size of the area and the number of discretized squares depend on the specific application domain. Figure 2 shows a basic scenario of the pheromone dynamics, focused to the most significant stages of diffusion and evaporation. The levels of pheromone intensity are represented by different grey gradations: the darker the gradation is, the higher the intensity.

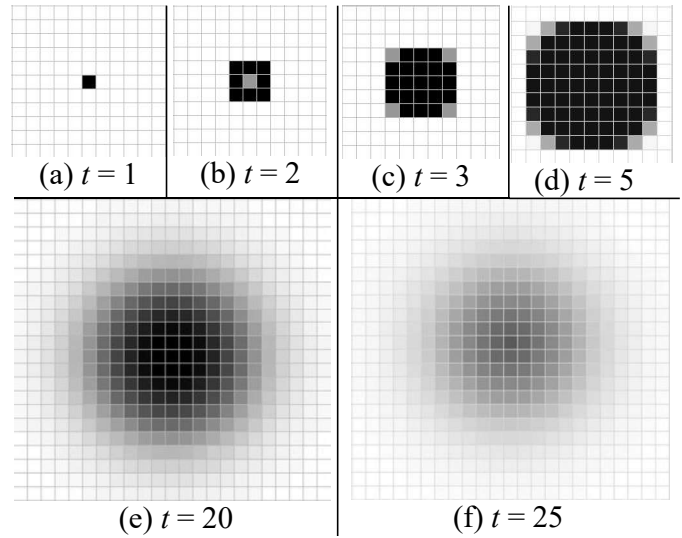

**Fig. 2.** Basic scenario of pheromone dynamics: (a) releasing; (b) mainly diffusing; (c-d) diffusing and evaporating; (e-f) mainly evaporating.

More specifically, in Fig. 2: (a) a single pheromone intensity *I* is released; (b) at the first steps, the pheromone is mainly diffusing (moving) to the nearby cells, with a constant diffusion rate $\delta \in [0,1]$, *StigDiffusion*; (c-d) the pheromone is diffusing and evaporating; by evaporating, the pheromone decreases its intensity over time; it is ruled by the constant rate $\varepsilon \in [0,1]$, *StigEvaporation*; (e-f) the pheromone is mainly evaporating. More formally, the pheromone intensity *p* released at the instant *t* on the cell *(x,y)* is then characterized by the

following dynamics:

$$p_{x,y}(t) = \varepsilon \cdot \left[ (1-\delta) \cdot p_{x,y}(t-1) + \Delta p_{x,y}(t-1,t) + d_{x,y}(t-1,t) \right] \qquad (1)$$

where $(1-\delta) \cdot p_{x,y}(t\text{-}1)$ represents the amount remaining after diffusion to nearby cells, $\Delta p_{x,y}(t\text{-}1,t)$ the additional deposits made within the interval $(t\text{-}1,t]$, and $d_{x,y}(t\text{-}1,t)$ the input pheromone diffused from all the nearby cells. The latter can be formally calculated as:

$$d_{x,y}(t-1,t) = \frac{\delta}{8} \sum_{\substack{i=-1 \\ (i,j)\neq(0,0)}}^{1} \sum_{j=-1}^{1} p_{x+i,y+j}(t-1) \qquad (2)$$

since each of the 8 neighbor cells propagates the portion $\delta$ of its pheromone to the cell $(x,y)$ at each update cycle. The total amount in (1) is also multiplied by $\varepsilon$ (*StigEvaporation*) to take the evaporation into account.

The Environment supports also the management of the target detection with imperfect sensors. It is assumed that each target sensing can provide both false positive and false negative. However, this occurs, with a certain error probability of $\xi$, only while checking the target cell or the cells adjacent to the target, as represented in Fig. 3.

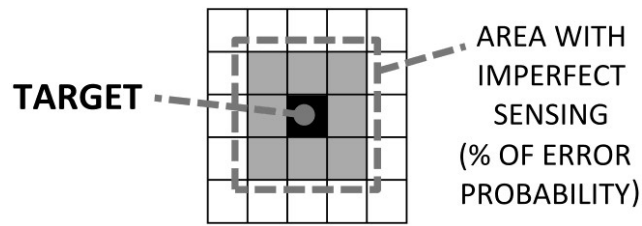

**Fig. 3.** Imperfect sensor model

In essence, it is introduced the notion of degradation of the sensing quality as a function of the proximity to the target: as the proximity increases, the sensing may generate an altered measure resulting in a wrong detection. This assumption implies that the discontinuities represented in Fig.1 should be small with respect to the source signal.

*4.2 The design of the drone behavior*

The drone behavior is structured into a prioritized logic, where each priority level implements one basic behavior, or role. At each update cycle, or tick, the role assumed by the drone is a consequence of the environmental sensing. In descending order or priority, the roles are: *obstacle avoider, tracker, flockmate* and *explorer*.

Figure 4 shows an overall representation of the drone behavior, using a UML (Unified Modeling Language) activity diagram. Here, every tick period, represented by the hourglass icon on bottom left, the environment updates his status, whereas the drone performs in parallel: the target detection, in which case it releases pheromone controlled by *StigDiffusion* and *StigEvaporationRate* parameters; the obstacle avoider. If a close object is detected, within the *ObstacleVision* radius, the drone points toward a free direction, when available, and moves forward. Otherwise, if there are no close objects detected, the drones play the *tracker* role: it tries to sense pheromone within the *Olfaction* radius and, if detected, points toward the pheromone peak. Alternatively, if pheromone is not detected, the drone plays the *flockmate* role: it tries to detect surrounding drones within the *FlockVision* radius, in order to point toward the flock. Finally, if there are no surrounding drones, as an *explorer* it performs a random turn within the *WiggleVar* angle, and then moves forward. Figure 5 represents a detailed modeling of the main procedures and roles played by a drone.

Figure 5a models the basic *drone behavior* consisting in releasing attractive pheromone with *StigIntensity* intensity, upon target detection and moving forward according to a given velocity set to *DroneVel*. In Fig. 5b, Fig. 5c, and Fig. 5d the *obstacle avoider*, the *tracker* and the *flockmate* roles are modelled, respectively. Figure 6 shows the main procedures of the flocking, according to a model called "Boids" in the literature [24]. In the flocking behavior, the drone takes into account only drones within a *FlockVision*. Figure 6a represents the *separation behavior*: drones close to others have to separate for better exploration; thus, if a drone senses another drone closer than the *MinimumSeparation*, it turns by an angle *MaxSeparateTurn*. Figure 6b shows the *alignment behavior*: the drone calculates the average direction of the drones in the flock vision and turns by an angle *MaxAlignTurn* to conform its direction to the flock direction. Figure 6c illustrates the *cohesion behavior*: the drone calculates the barycenter of the drones in the Flock vision and turns by an angle *MaxCohereTurn* towards the barycenter.

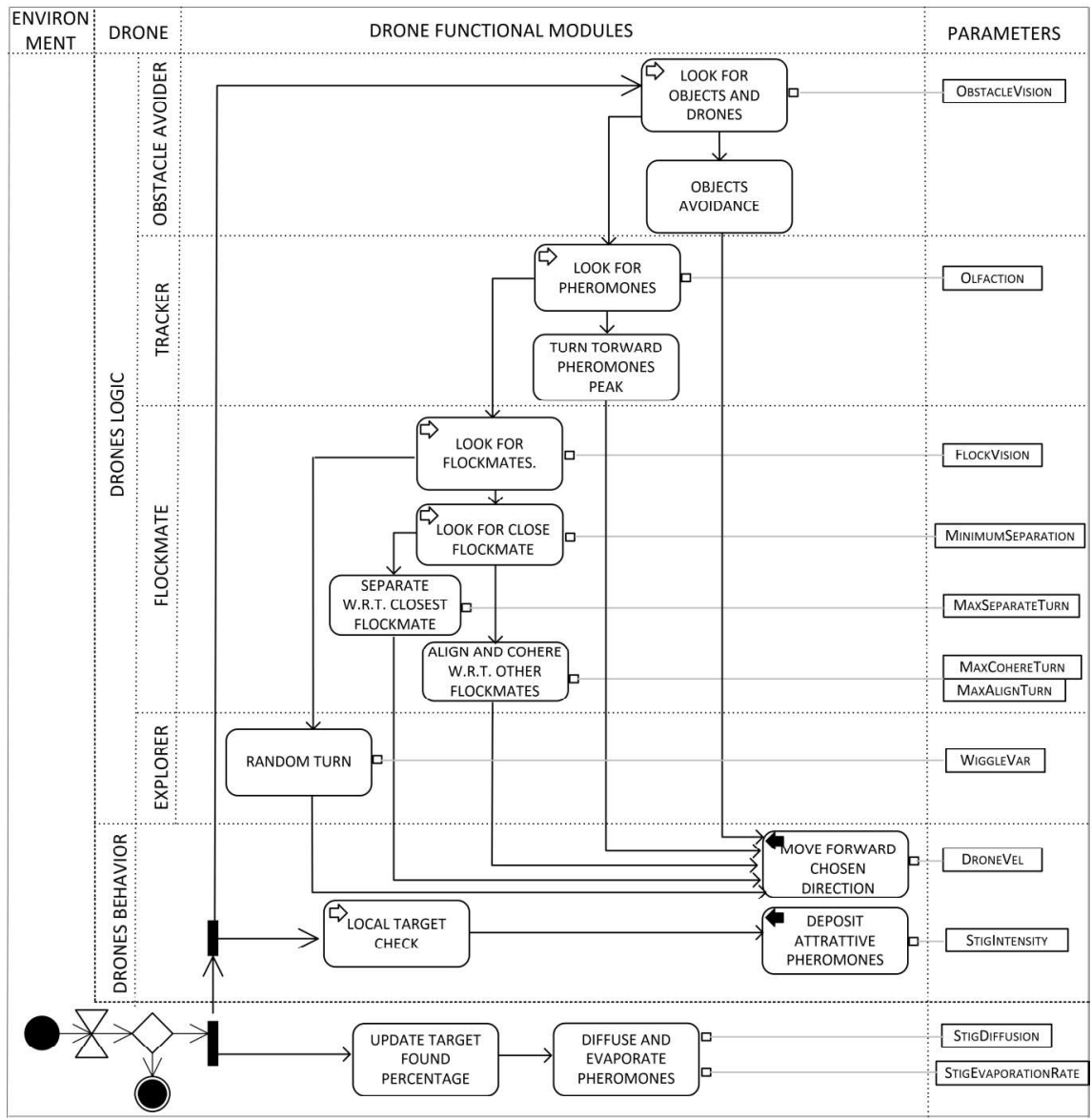

**Fig. 4.** Overall modeling of the drone behavior modularized in roles.

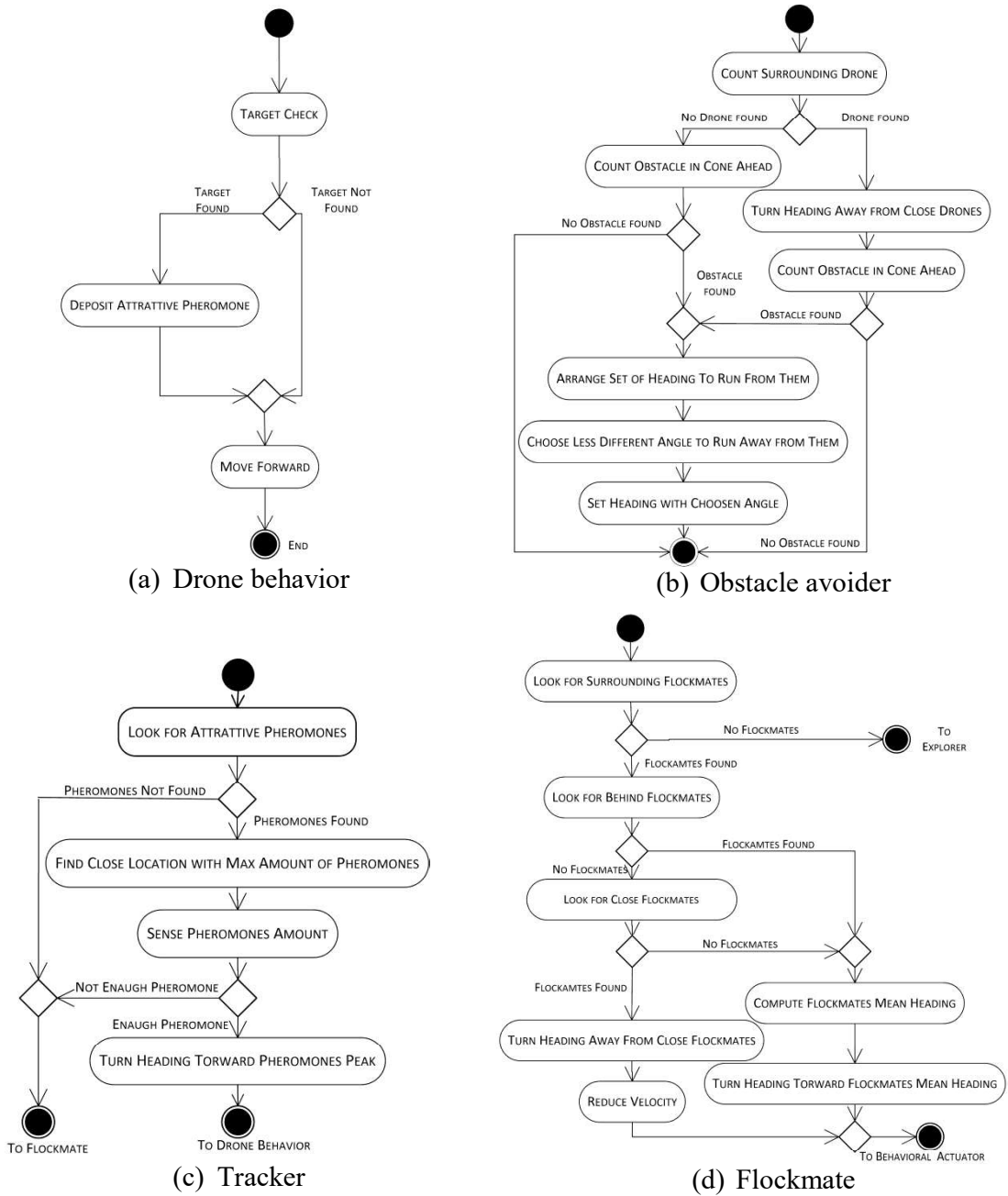

(a) Drone behavior

(b) Obstacle avoider

(c) Tracker

(d) Flockmate

**Fig. 5.** Detailed modeling of the main roles played by a drone

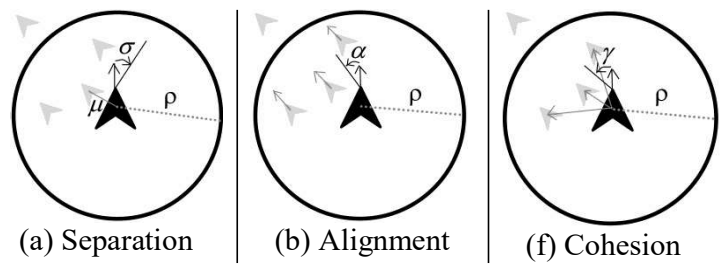

(a) Separation  (b) Alignment  (f) Cohesion

**Fig. 6.** Illustration of the procedures of the flocking behavior

## 6 Experimental studies

The proposed model has been implemented on *NetLogo*[1], a leading simulation platform for swarm intelligence. The output of the system is the total time needed to find the 95% of targets. According to Fig. 4, the model requires 12 parameters, to be tuned via three-phases: early analysis, under the assumption of reliable sensing (that is, *sensing error probability* and *sensing redundancy* set to 0.1 and 1, respectively); *parameter sensitivity analysis* on representative scenarios, by evaluating the uncertainty in the output for each parameter; finally, accurate setting on each of the most sensitive parameters, via a bisection method to find the value minimizing the output. For the reader's convenience, Table 1 summarizes the main structural and behavioral parameters of the model, with their range and their value set.

**Table 1** Structural and behavioral parameters.

| Name | Description (unit measure) | Range | Set v. |
|------|----------------------------|-------|--------|
| DroneVel | Drone horizontal speed (m/s) | (0,15) | 1 |
| WiggleVar | Drone max rand-fly turn angle (°) | (0,180) | 150 |
| ObstacleVision | Drone object sensing distance (m) | (0, 5) | 2 |
| FlockVision | Flock visibility radius (m) | [0, 50] | 7 |
| MinimumSeparation | Flock mobility distance (m) | [0,5] | 3 |
| MaxSeparateTurn | Flock separation angle (°) | (0,180) | 30 |
| MaxAlignTurn | Flock alignment angle (°) | (0,180) | 20 |
| MaxCohereTurn | Flock cohesion angle (°) | (0,180) | 5 |
| Olfaction | Pheromone sensing distance (m) | $(0, \infty)$ | 1 |
| StigIntensity | Pheromone release intensity | $(0, \infty)$ | 40K |
| StigDiffusion | Pheromone diffusion rate (%) | [0,1] | 0.85 |
| StigEvaporation | Pheromone evaporation rate (%) | [0,1] | 0.05 |
| SensingError | Sensing error probability (%) | (0, 100) | [0.1,1] |
| Redundancy | Sensing Redundancy | $(0, \infty)$ | {1,3,5} |

The algorithm has been tested on four different scenarios, such as *Field, Dumps, Urban* and *Mines*. The *Field* scenario is made by 5 groups of targets scattered over the area, with about 10 targets per group. There are no obstacles. Figure 7 shows a snapshot with the spatial arrangement of flocks of different forms and sizes, together with four stigmergic formations on corresponding groups of targets. Here, it can be observed that stigmergic formations attracted flocks of drones.

---

[1] https://ccl.northwestern.edu/netlogo/

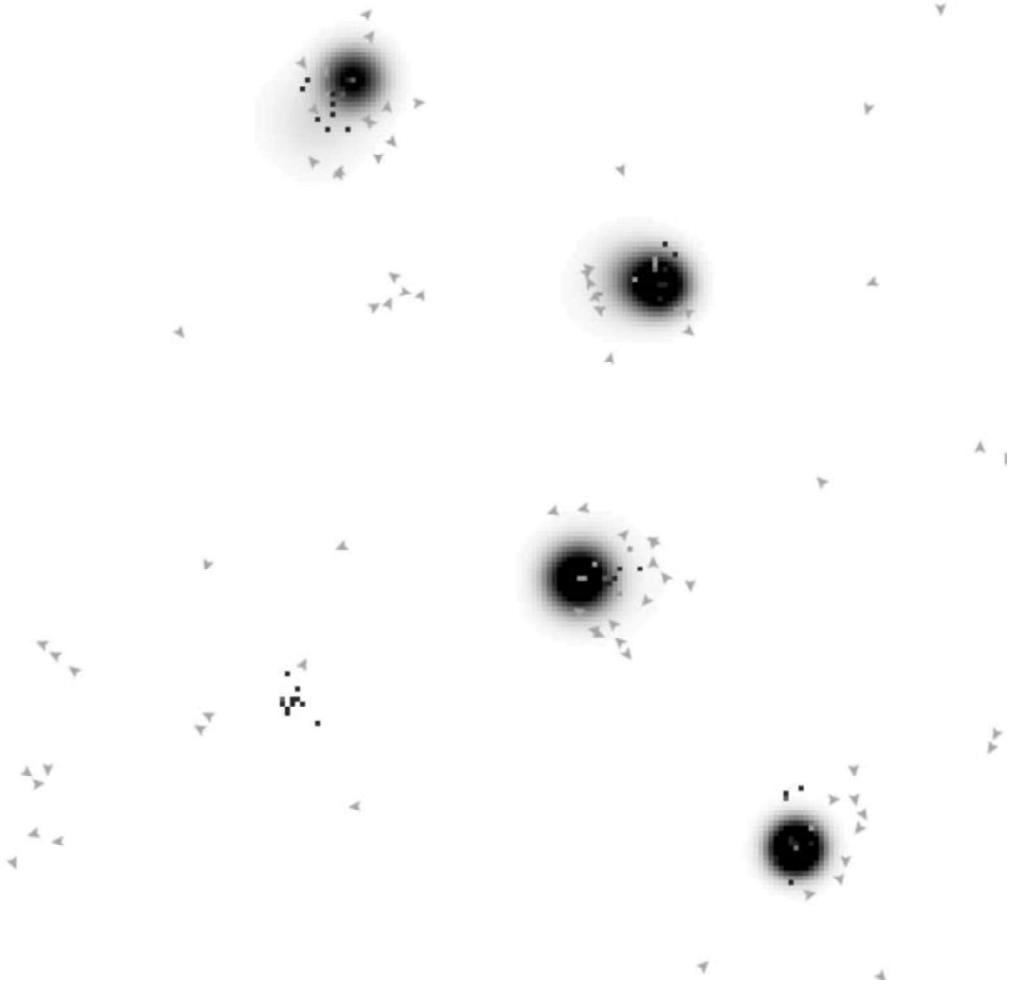

**Fig. 7.** A snapshot of the *Field* scenario with flocks and stigmergic formations

An initial configuration of the Field scenario is shown in Fig. 8a. Here, 80 total drones (represented by triangular forms) are arranged into four dense flocks, placed at the antipodes of the area, whereas the targets are represented by clusters of black dots. The second scenario, called *Dumps* (Fig. 8b) represents a synthetic reconstruction of woodland with three abusive garbage aggregations, modelled by three groups of targets. Here, 30 targets and 100 trees are represented by gray and black dots, respectively. 80 total drones, arranged into 4 flocks, are initially placed at the antipodes of the area. The third scenario, that is, *Urban* (Fig. 8c) is characterized by two cluster of 110 total targets placed on two sides of corresponding buildings. Overall, 7 total buildings are located. 40 drones, arranged into four flocks, are placed at the antipodes of the area, with no trees at all. Finally, the *Urban Mines* (Fig. 8d) scenario is derived from real-world examples of areas near Sarajevo, in Bosnia-Herzegovina, with landmine objects, selected from publicly available maps[2]. Recently, some authors actually proposed the use of mini-UAVfor detecting landmines [34]. Drones have been initially placed on the boundaries of the area. With respect to the map of the first three scenarios, whose area is 200 square meters, in the last scenario the area is 400 square meters.

To carry out the experiments under the requirement of imperfect sensor model, a sensing error probability in the interval [0.1, 1] percent with uniform distribution has been added. Then, the system output has been evaluated by requiring a prefixed number of repeated measures of the targets in the termination criterion, that is, sensing redundancy values 3 and 5.

To assess the effectiveness of the proposed approach, the performance of the model has been evaluated on three approaches: Random Fly ("R"), Stigmergic approach ("S"), Stigmergic and Flocking aproach ("S+F"). For each experiment, 10 trials have been carried out. It has been determined that the resulting performance indicator samples are well-modeled by a normal distribution, using a graphical normality test. Hence, the 95% confidence intervals have been calculated. Table 2 summarizes, for each scenario, the characteristics and the results in the form "mean ± confidence interval". The results confirm that the use of stigmergy speeds up the target search process in any scenario. Moreover, results become even better in combination with flocking. It can be remarked that all scenarios have been processed by using a general purpose parameterization determined a with reliable sensor model. Indeed, a parameterization ad initialization adapted to types of scenario might produce better results.

---

[2] http://www.see-demining.org/main.htm

**Table 2**. Characteristics and numerical results (mean ± confidence interval) of each scenario.

| | | Field | Dumps | Urban | Urban Mines |
|---|---|---|---|---|---|
| | # targets | 50 | 30 | 110 | 40 |
| | # clusters | 5 | 3 | 2 | 40 |
| | # trees | 0 | 100 | 0 | 54 |
| | # buildings | 0 | 0 | 7 | 28 |
| | # drones | 80 | 80 | 40 | 25 |
| **Algorithm (redundancy)** | **R (1)** | 2,604 ± 248 | 2,252 ± 212 | 2,340 ± 229 | 651 ± 55 |
| | **S (1)** | 1,383 ± 126 | 1,297 ± 102 | 1,748 ± 188 | 560 ± 49 |
| | **S+F (1)** | 1,078 ± 106 | 1,009 ± 141 | 1,259 ± 102 | 487 ± 29 |
| | **R (3)** | 4,161 ± 269 | 3,993 ± 266 | 3,688 ± 286 | 944 ± 55 |
| | **S (3)** | 1,758 ± 151 | 1,513 ± 116 | 2,089 ± 197 | 707 ± 84 |
| | **S+F (3)** | 1,484 ± 147 | 1,289 ± 135 | 1,861 ± 166 | 594 ± 34 |
| | **R (5)** | 6,173 ± 361 | 6,163 ± 399 | 4,647 ± 271 | 1,167 ± 51 |
| | **S (5)** | 2,109 ± 246 | 2,208 ± 208 | 2,488 ± 280 | 770 ± 93 |
| | **S+F (5)** | 1,591 ± 136 | 1,823 ± 233 | 2,102 ± 151 | 726 ± 32 |

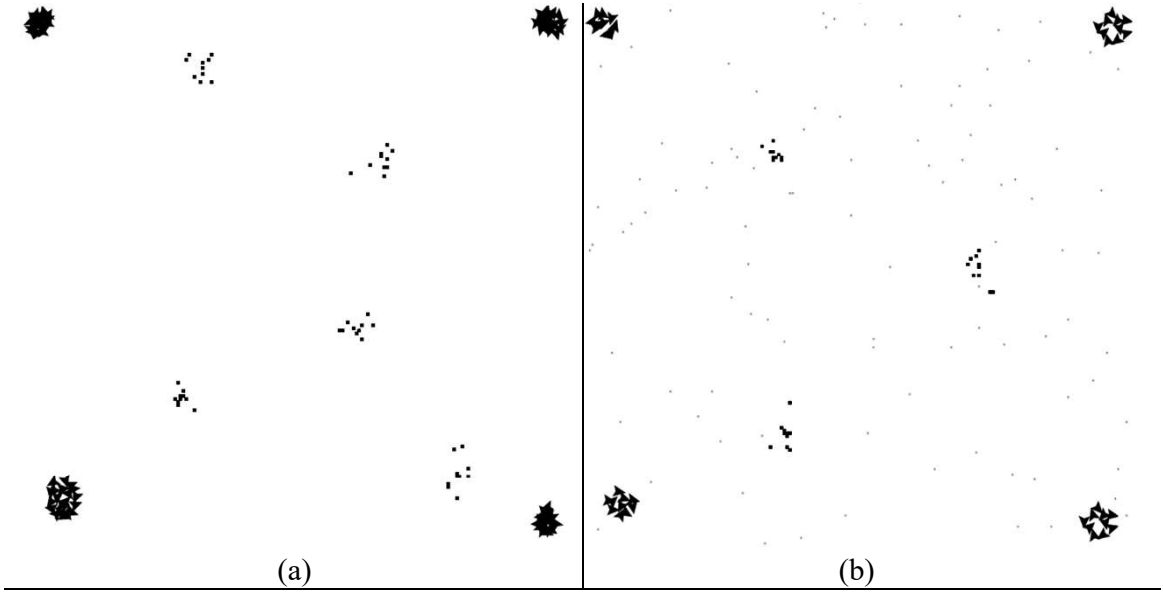

(a)                              (b)

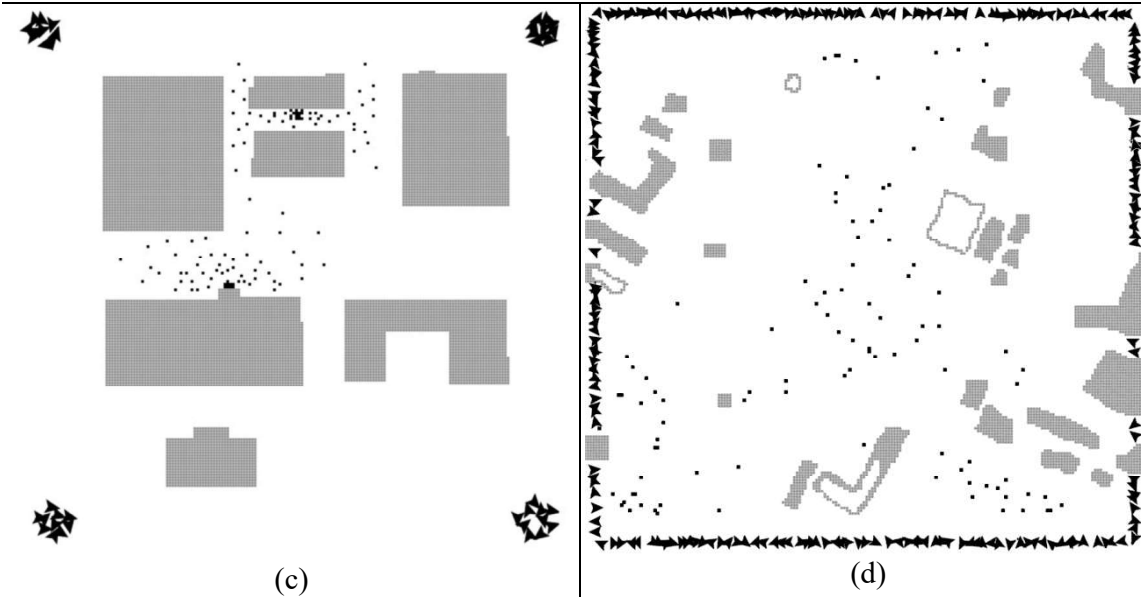

<div align="center">(c)          (d)</div>

**Fig. 8.** Models of three synthetic and one real-world scenarios: (a) Field; (b) Dumps; (c) Urban; (d) Urban Mines.

To better highlight the scalability of our approach against redundancy, Fig. 9a-d shows the completion time for redundancy 1, 3 and 5, for each scenario. Here, it is apparent that Stigmergy introduces a significant improvement of trend over Random Fly, both alone and combined with flocking behavior.

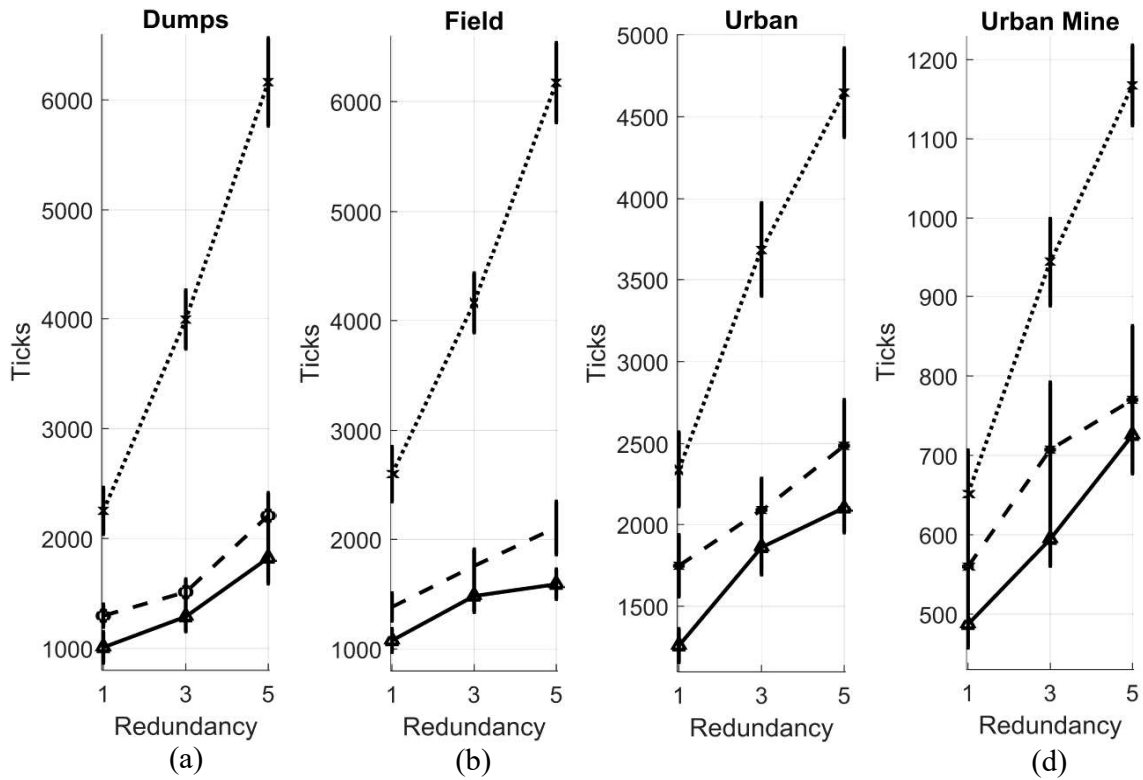

(c)

**Fig. 9.** Completion time against redundancy, for each scenario and with different approaches: Random Fly (dotted line), Stigmergy (dashed), and Stigmergy + Flocking (solid).

## 7 Conclusions and Future Work

In this paper, it is presented a novel swarm approach for coordinating mini-UAVs to perform target search with imperfect sensors. The approach uses a stigmergic behavior, consisting in the release of information in the environment in the form of attractive digital pheromones, in areas where potential targets are sensed. Moreover, the approach employs flocking behavior, resulting in a flexible arrangement of drones according to the stigmergic potential field. The paper illustrates the approach from the behavioral and architectural point of views, and then discusses the experimental studies. Results on synthetic and real-world scenarios prove the benefits of both stigmergy and flocking, in terms of tolerance to errors and scalability for increasing redundancy requirements.

The overall mechanism can be better enabled if structural parameters are correctly tuned for the given scenario. Determining such correct parameters is not a simple task since different areas have different features. Thus, an appropriate tuning to adapt parameters to the specific search area is desirable to make the search more effective. For this purpose, to use a parameter optimization strategy is considered a key investigation activity for future work.

## Acknowledgements

This work was carried out in the framework of the SCIADRO project, cofunded by the Tuscany Region (Italy) under the Regional Implementation Programme for Underutilized Areas Fund (PAR FAS 2007-2013) and the Research Facilitation Fund (FAR) of the Ministry of Education, University and Research (MIUR)

**References**

1.  K. Whitehead, C. H. Hugenholtz, S. Myshak, O. Brown, A. LeClair, A. Tamminga, T. E. Barchyn, B. Moorman, B. Eaton, Remote sensing of the environment with small unmanned aircraft systems (UASs), part 2: scientific and commercial applications, *Journal of Unmanned Vehicle Systems*, **02** (2014), 86-102.

2.  A. Errico, C. V. Angelino, L. Cicala, G. Persechino, C. Ferrara, M. Lega, A. Vallario, C. Parente, G. Masi, R. Gaetano, G. Scarpa, D. Amitrano, G. Ruello, L. Verdoliva, G. Poggi, Detection of environmental hazards through the feature-based fusion of optical and SAR data: a case study in southern Italy, *International Journal of Remote Sensing*, **36** (2015), 3345-3367.

3.  A. Errico, C. V. Angelino, L. Cicala, D. P. Podobinski, G. Persechino, C. Ferrara, M. Lega, A. Vallario, C. Parente, G. Masi, R. Gaetano, G. Scarpa, D. Amitrano, G. Ruello, L. Verdoliva, G. Poggi, SAR/multispectral image fusion for the detection of environmental hazards with a GIS, *Proceedings of SPIE - The International Society for Optical Engineering*, **9245** (2014), doi: 10.1117/12.2066476.

4.  M. Lega, R. M. A. Napoli, A new approach to solid waste landfills aerial monitoring, *WIT Transactions on Ecology and the Environment*, **109** (2008), 193-199.

5.  M. Lega, R. M. A. Napoli, Aerial infrared thermography in the surface waters contamination monitoring, *Desalination and Water Treatment*, **23** (2010), 141-151.

6.  G. Persechino, P. Schiano, M. Lega, R. M. A. Napoli, C. Ferrara, J. Kosmatka, Aerospace-based support systems and interoperability: The solution to fight illegal dumping, *WIT Transactions on Ecology and the Environment*, **140** (2010), 203-214.

7.  G. Persechino, M. Lega, G. Romano, F. Gargiulo, L. Cicala, IDES project: An advanced tool to investigate illegal dumping, *WIT Transactions on Ecology and the Environment*, **173** (2013), 603-614.

8.  M. Lega, C. Ferrara, G. Persechino, P. Bishop, Remote sensing in environmental police investigations: Aerial platforms and an innovative application of thermography to detect several illegal activities, *Environmental Monitoring and Assessment*, **186** (2014), 8291-8301.

9.  M. Lega, L. D'Antonio, R. M. A. Napoli, Cultural Heritage and Waste Heritage: Advanced techniques to preserve cultural heritage, exploring just in time the ruins produced by disasters and natural calamities, *WIT Transactions on Ecology and the Environment*, **140** (2010), 123-134.

10. M. Lega, G. Persechino, GIS and infrared aerial view: Advanced tools for the early detection of environmental violations, *WIT Transactions on Ecology and the Environment*, **180** (2014), 225-235.

11. F. Gargiulo, G. Persechino, M. Lega, A. Errico, IDES project: A new effective tool for safety and security in the environment, *Lecture Notes in Computer Science (LNCS)*, **8286** (2013), 201-208.

12. M. Lega, D. Ceglie, G. Persechino, C. Ferrara, R. M. A. Napoli, Illegal dumping investigation: A new challenge for forensic environmental engineering, *WIT Transactions on Ecology and the Environment*, **163** (2012), 3-11.

13. M. Lega, J. Kosmatka, C. Ferrara, F. Russo, R. M. A. Napoli, G. Persechino, Using Advanced Aerial Platforms and Infrared Thermography to Track Environmental Contamination, *Environmental Forensics*, **13** (2012), 332-338.

14. L. F. Bertuccelli, J. P. How, Robust UAV search for environments with imprecise probability maps, *Proceedings of the 44th IEEE Conference on Decision and Control*, (2005), 5680-5685.

15. B. Bethke, M. Valenti, J. How, Cooperative vision based estimation and tracking using multiple UAVs, In *Advances in Cooperative Control and Optimization*, Springer, Berlin Heidelberg, (2007), 179-189.

16. R. McCune, R. Purta, M. Dobski, A. Jaworski, G. Madey, Y. Wei, A. Madey and M.B. Blake, Investigations of DDDAS for command and control of uav swarms with agent-based modeling, *Proceedings of the 2013 Winter Simulation Conference: Simulation: Making Decisions in a Complex World*, (2013), 1467-1478.

17. M. Brambilla, E. Ferrante, M. Birattari, M. Dorigo, Swarm robotics: a review from the swarm engineering perspective, *Swarm Intelligence*, **7** (2013), 1-41.

18. J. A. Sauter, R. Matthews, H. Van Dyke Parunak, S. A. Brueckner, Performance of digital pheromones for swarming vehicle control, *Proceedings of the fourth international joint conference on Autonomous agents and multiagent systems,* (2005), 903-910.

19. M.G.C.A. Cimino, A. Lazzeri, G. Vaglini, Combining stigmergic and flocking behaviors to coordinate swarms of drones performing target search, *Proceedings of the Sixth International Conference on Information, Intelligence, Systems and Applications* (2015), 1-6.

20. H. Van Dyke Parunak, M. Purcell, R. O'Connell, Digital pheromones for autonomous coordination of swarming UAV's, *Proceedings of the 1st Technical Conference and Workshop on Unmanned Aerospace Vehicles*, (2002), 48105-1579.

21. G. Ermacora, A. Toma, B. Bona, M. Chiaberge, M. Silvagni, A cloud robotics architecture for an emergency management and monitoring service in a smart city environment, *Proceedings of the International Conference of Intelligent Robots and Systems*, (2013).

22. S. Reece, S. Roberts, C. Claxton, D. S. Nicholson, Multi-sensor fault recovery in the presence of known and unknown fault types, *Proceedings of the 12th International Conference on Information Fusion*, (2009).

23. A. F. T. Winfield, C. J. Harper and J. Nembrini, Towards the Application of Swarm Intelligence in Safety Critical Systems, *Proceedings of the International Conference on System Safety,* (2006), 89-95.

24. C. W. Reynolds, Flocks, herds and schools: A distributed behavioral model*, Proceedings of the ACM Siggraph Conference on Computer Graphics*, (1987), 25-34 .

25. C. Bernon, M. P. Gleizes, G. Picard, Enhancing self-organising emergent systems design with simulation, *Engineering Societies in the Agents World VII*, Springer Berlin Heidelberg, (2007), 284-299.

26. T. Kuyucu, I. Tanev, K. Shimohara, Superadditive effect of multi-robot coordination in the exploration of unknown environments via stigmergy, *Neurocomputing*, **148** (2015), 83-90.

27. P. Dasgupta, A multiagent swarming system for distributed automatic target recognition using unmanned aerial vehicles, *IEEE Transactions on Systems, Man, and Cybernetics—Part A: Systems and Humans*, **38** (2008) 549-563.

28. Y. Tan,  Z. Zhong-Yang, Research advance in swarm robotics, *Defence Technology* **9** (2013), 18-39.

29. N. Bouraqadi, A. Doniec, E. M. de Douai. Flocking-based multi-robot exploration. *Proceedings of National conference on control architectures of robots*, (2009).

30. S. Hauert, S. Leven, M. Varga, F. Ruini, A. Cangelosi, J.-C. Zufferey, D. Floreano, Reynolds flocking in reality with fixed-wing robots: communication range vs.

maximum turning rate, *Proceedings of the International Conference on Intelligent Robots and Systems*, (2011).

31. P. Bresciani, A. Perini, P. Giorgini, F. Giunchiglia, J. Mylopoulos, Tropos: An agent-oriented software development methodology, *Autonomous Agents and Multi-Agent Systems*, **8** (2004) , 203–236.

32. F. Pagliarecci, L. Penserini, L. Spalazzi, From a Goal-Oriented methodology to a BDI agent language: the case of Tropos and Alan, *Proceedings of Workshop on Agents, Web Services and Ontologies Merging*, (2007).

33. P. Giorgini, J. Mylopoulos, A. Perini, A. Susi, The Tropos Metamodel and its Use, *Informatica*, **29** (2005), 401.408.

34. J. Rodriguez, C. Castiblanco, I. Mondragon, J. Colorado, Low-cost quadrotor applied for visual detection of landmine-like objects, *Proceedings of the International Conference on Unmanned Aircraft Systems*, (2014).